\documentclass[12pt]{article}

\usepackage[dvips]{graphicx}
\usepackage{cite}

\def\tlab{\mbox{$T_{\rm Lab}$}}

\def\lamr{\lambda^r}
\def\lami{\lambda^i}

\def\pzp{p_0^{\prime}}

\newcommand{\bp}{{\bf p}}
\newcommand{\bpp}{{\bf p}^{\prime}}
\newcommand{\bk}{{\bf k}}

\def\mth{m_{th}}

\begin{document}

\begin{center}
{\bfseries
RELATIVISTIC SEPARABLE INTERACTION KERNEL OF THE NEUTRON-PROTON SYSTEM
WITH INELASTICITIES}

\vskip 5mm

S.G. Bondarenko$^{1 \dag}$, V.V. Burov$^{1}$, E.P. Rogochaya$^{2}$

\vskip 5mm

{\small
(1) {\it
Bogoliubov Laboratory of Theoretical Physics,\\
Joint Institute for Nuclear Research, Dubna, Russia
}
\\
(2) {\it
Veksler and Baldin Laboratory of High Energy Physics,\\
Joint Institute for Nuclear Research, Dubna, Russia
}
\\
$\dag$ {\it
E-mail: bondarenko@jinr.ru
}}
\end{center}

\vskip 5mm

\begin{center}
\begin{minipage}{150mm}
\centerline{\bf Abstract}
Within a covariant Bethe-Salpeter approach,
the relativistic complex separable neutron-proton interaction
kernel is proposed. The uncoupled partial-wave states
with the total angular momentum $J$=0,1 are considered.
The multirank separable potentials elaborated earlier are real-valued and,
therefore, enable to describe the elastic part (phase shifts, low-energy
parameters, etc.) of the scattering only.
The description of the inelasticity parameter comes out of the imaginary
part introduced into them. To obtain parameters
of the complex potentials the elastic neutron-proton
scattering experimental data up to 3 GeV are used. A signal of dybaryon resonances
in the $^3P_0^+$ partial-wave state is discussed.
\end{minipage}
\end{center}

\vskip 10mm

\section{Introduction}\label{sect1}

Using the Bethe-Salpeter (BS) equation~\cite{Salpeter:1951sz}
to describe the nucleon-nucleon (NN) interaction is
one of the most consistent approaches. In this
formalism, one has to deal with a system of nontrivial integral equations
for the NN scattered states and for the bound state -- the deuteron.
To solve the system of integral equations it is convenient to use
a separable ansatz~\cite{Bondarenko:2002zz} for the interaction kernel
in the BS equation. In this case, one can
transform integral equations into a system of algebraic linear ones which
is easy to solve.
Parameters of the interaction kernel are found from the analysis of phase
shifts for respective partial-wave states and low-energy parameters
as well as deuteron static properties
(bound state energy, magnetic moment, etc.).

In our previous papers~\cite{npa1,npa2} the multirank separable potentials
for the description of the scattered neutron-proton ($np$) system
with the total angular momentum
$J=0,1$ and the bound state -- the deuteron -- were proposed.
Various methods of a relativistic generalization of initially
nonrelativistic separable functions parametrizing the interaction kernel
were considered.
The elaborated potentials allow us to describe the experimental data
for the phase shifts up to the laboratory kinetic energy $\tlab \sim 3$ GeV,
static properties of the deuteron, and the exclusive electron-deuteron
breakup in the plane-wave approximation~\cite{npa1,npa2,fbs}.

However, it is well known that the influence of the inelastic channels concerned with
non-nucleon degrees of freedom (mesons, $\Delta$ isobars, nucleon excitations,
six-quark admixtures, etc.) becomes significant with increasing
energy of the nucleon-nucleon system. To treat them in the elastic NN
scattering the inelasticity parameter which is responsible
for a proper flux behavior is introduced.

There are several methods to describe the inelasticity parameter (see e.g.,
\cite{eyser,knyr}). One of the way is to use a complex
NN potential instead of the real-valued one. We apply this idea to the
relativistic separable interaction
kernel obtained earlier~\cite{npa1,npa2}.
At the same time we want to keep the results
for observables below the inelasticity threshold and have a slight
difference above it. To achieve this we consider the complex separable
interaction kernel of a special type (Sec.\ref{sect3}).
A special procedure which we apply
to find new imaginary interaction kernel parameters is described
in Sec.\ref{sect4}.
The discussion and conclusion are given in Sec.\ref{sect5}
and Sec.\ref{sect6}, respectively.

\section{Parametrization of the S matrix}\label{sect2}
In the paper, we use the Arndt-Roper parametrization~\cite{arndtroper}
of the elastic NN scattering $S$ matrix. For uncoupled partial-wave states,
in the presence of inelasticity the $S$ matrix is written via the $K$ matrix
as follows:
\begin{eqnarray}
S = \frac{1-K_i+iK_r}{1+K_i-iK_r}=\eta\, \exp(2i\delta),
\label{smparam}
\end{eqnarray}
where real $K_r$ and imaginary $K_i$ parts of the $K$ matrix
($K = K_r + i K_i$) are parametrized
\begin{eqnarray}
K_r = \tan\delta, \quad K_i = \tan^2\rho,
\label{eta1}
\end{eqnarray}
in terms the phase shift $\delta$ and the inelasticity parameter $\rho$, respectively, and
\begin{eqnarray}
\eta^2 = \frac{1+K^2-2K_i}{1+K^2+2K_i},\hskip 11mm\nonumber\\
K^2=K_r^2+K_i^2.\hskip 20mm
\label{eta2}
\end{eqnarray}
%\begin{eqnarray}
%\delta=\frac12\{\tan^{-1}[K_r/(1-K_i)]+\tan^{-1}[K_r/(1+K_i)]\}.
%\label{delta}
%\end{eqnarray}
For elastic scattering ($\rho=0$), $\delta=\delta_e$, $\eta=1$ and
$S=S_e=\exp(2i\delta_e)$.

\section{Complex separable kernel}\label{sect3}
We assume that the interaction kernel $V$ conserves parity,
the total angular momentum  $J$ and its projection, and isotopic spin.
Due to the tensor nuclear force, the orbital angular momentum
$L$ is not conserved. The negative-energy two-nucleon
states are switched off, which leads to the total spin $S$
conservation. The partial-wave-decomposed BS equation is, therefore,
reduced to the following form:
\begin{eqnarray}
T_{l'l}(\pzp, |\bpp|; p_0, |\bp|; s) =
V_{l'l}(\pzp, |\bpp|; p_0, |\bp|; s)
\hskip 50mm
\label{BS}\\
+ \frac{i}{4\pi^3}\sum_{l''}\int\limits_{-\infty}^{+\infty}\!
dk_0\int\limits_0^\infty\! \bk^2 d|\bk|\, \frac{V_{l'l''}(\pzp,
|\bpp|; k_0,|\bk|; s)\, T_{l''l}(k_0,|\bk|;p_0,|\bp|;s)}
{(\sqrt{s}/2-E_{\bk}+i\epsilon)^2-k_0^2}, \nonumber
\end{eqnarray}
where $l=l^{\prime}=l^{\prime\prime}$ for spin-singlet
and uncoupled spin-triplet states. The square of the $np$ pair total momentum
$s$ is connected with the laboratory energy $\tlab$ as:
$s=2m\tlab+4m^2$, $m$ is the mass of the nucleon.

To describe the inelasticity in the
elastic NN scattering we modify the real-valued relativistic potential
adding the imaginary part:
$$ V_r \to V = V_r + iV_i.$$

To solve the Eq.(\ref{BS}) the separable (rank $N$) ansatz~\cite{Bondarenko:2002zz}
for the NN interaction kernel is used:
\begin{eqnarray}
V_{l'l}(\pzp, |\bpp|; p_0, |\bp|; s)=\sum_{m,n=1}^N
\Big[\lamr_{mn}(s) + i \lami_{mn}(s)\Big]
g_i^{[l']}(\pzp, |\bpp|)g_j^{[l]}(p_0, |\bp|), \label{V_separ}
\end{eqnarray}
where the imaginary part $\lami$ has the form:
\begin{eqnarray}
\lami_{mn}(s) = \theta(s-s_{th})\, \Big(1-\frac{s_{th}}{s}\Big)\,{\bar\lambda}^i_{mn},
\label{lami}
\end{eqnarray}
$g_j^{[l]}$ are the model functions, $\lambda_{mn} = \lambda^{r}_{mn}+i\lambda^{i}_{mn}$
is a matrix of model parameters and $s_{th}$ is
the inelasticity threshold (the first energy point where the inelasticity
becomes nonzero). In this case, the resulting $T$ matrix has a similar
separable form:
\begin{eqnarray}
T_{l'l}(\pzp, |\bpp|; p_0, |\bp|; s)=
\sum_{m,n=1}^N\tau_{mn}(s)g_i^{[l']}(\pzp, |\bpp|)
g_j^{[l]}(p_0, |\bp|),
\end{eqnarray}
where
\begin{eqnarray}
\big(\tau_{mn}(s)\big)^{-1} = \Big(\lamr_{mn}(s)+i\lami_{mn}(s)\Big)^{-1}+h_{mn}(s),
\end{eqnarray}
\begin{eqnarray}
h_{mn}(s)=-\frac{i}{4\pi^3}\sum_{l}\int dk_0\int
\bk^2d|\bk| \frac{g_m^{[l]}(k_0,|\bk|)g_n^{[l]}(k_0,|\bk|)}{(\sqrt
s/2-E_{\bk}+i\epsilon)^2-k_0^2}. \label{H_separ}
\end{eqnarray}
It should be noted that functions $g_m^{[l]}$ and parameters
$\lamr$ coincide with those used in~\cite{npa1,npa2}
while $\lami$ are new parameters which are calculated.

The separable functions $g_m^{[l]}$ used in the representation (\ref{V_separ})
of the interaction kernel $V$ are obtained by a relativistic
generalization of initially nonrelativistic Yamaguchi-type
functions depending on the 3-momentum squared $|\bp|$.
We introduce the imaginary part $V_i$ of the potential $V$ (\ref{V_separ})
adding the new parameters $\lami$ to the real part $V_r$ which is left intact.
Thus, we intend to describe the additional inelasticity parameters by a minimal
change of the previous kernels~\cite{npa1,npa2}.

\section{Calculations and results}\label{sect4}
We start from the real-valued interaction kernels which were obtained from
the minimization of the squared derivative function $\chi^2$ containing
the phase shifts and the
low-energy characteristics (details can be found in~\cite{npa1,npa2}).
Then we fix the parameters of the real part ($\lamr$, $\beta$ and $\alpha$)
and calculate the parameters $\lami$ to describe the inelasticity.

The calculation is performed for all available experimental data
for the phase shifts and the inelasticity parameters taken from the SAID
program~\cite{said}.

The minimization procedure for the function
\begin{eqnarray}
\chi^2=
\sum\limits_{m=\mth}^{n} (\delta^{\rm exp}(s_m)-\delta(s_m))^2/(\Delta\delta^{\rm exp}(s_m))^2+
\sum\limits_{m=\mth}^{n} (\rho^{\rm exp}(s_m)-\rho(s_m))^2/(\Delta\rho^{\rm exp}(s_m))^2
\label{mini_p}
\end{eqnarray}
is used for every partial-wave state. Here $n$ is a number of available experimental points.
The number $\mth$
corresponds to the data point with the first nonzero $\rho$ value.
It is defined by the threshold kinetic energy $\tlab_{\,th}$ which is taken from the
single-energy analysis~\cite{said}.

Thus, given the real part of the separable potential the imaginary part
parameters $\lami$ enable to describe the inelasticity with a minimal change
of the phase shift description.

The obtained parameters $\lami$ are listed in Tables~\ref{p_param} and~\ref{1s0_param}.

In Figs.\ref{1p1}-\ref{1s0}, the results of the phase
shift and inelasticity parameter calculations (MYI2, MYI3 - red dashed line)
are compared with the experimental data,
our previous result without inelasticities~\cite{npa1,npa2}
(MY2, MY3 - red solid line; only for phase shifts)
and the SP07 solution~\cite{Arndt:2007qn} (green dashed-dotted line).

\begin{center}
\begin{table}
\caption{Parameters $\bar\lami$ of the rank-two kernel for $P$ partial-wave states.
(We would like to note mispints in Table 1~\cite{npa1} where $\bar\lambda$ (GeV$^4$)
should be read as $\bar\lambda$ (GeV$^2$) for the $^3P_0^+$ partial-wave state).}
\centering
\begin{tabular}{l|ccc}
\hline\hline
                                &               & MYI2      &              \\
\hline
                                &$^1P_1^+$      & $^3P_0^+$  & $^3P_1^+$     \\
\hline\\[-4mm]
$\bar\lami_{11}$    (GeV$^4$) & -0.007097474  &  97.12885  & -0.007617132  \\
$\bar\lami_{12}$    (GeV$^4$) & -0.692547     & -114.857   & -0.3582908    \\
$\bar\lami_{22}$    (GeV$^4$) & -67.62616     & -35.16663  & -11.10021     \\
$\tlab_{\,th}$        (GeV$$)   &  0.35         &  0.25      &  0.35         \\
\hline\hline
\end{tabular}\label{p_param}
\end{table}
\end{center}
\begin{center}
\begin{table}
\caption{Parameters $\bar\lami$ of the rank-three kernel for the $^1S_0^+$ state.
(We would like to note mispints in Table 2~\cite{npa1} where $\bar\lambda$ (GeV$^2$
should be read as $\bar\lambda$ (GeV$^0$)).}
\centering
\begin{tabular}{lcclc}
\hline\hline

                               &  MYI3          \\
\hline\\[-4mm]
$\bar\lami_{11}$    (GeV$^2$) &  -0.01332595   \\
$\bar\lami_{12}$    (GeV$^2$) &  -89.63644     \\
$\bar\lami_{13}$    (GeV$^2$) &    0.151908    \\
$\bar\lami_{22}$    (GeV$^2$) &  -58097.6      \\
$\bar\lami_{23}$    (GeV$^2$) &   2276.805     \\
$\bar\lami_{33}$    (GeV$^2$) &  -217.6001     \\
$\tlab_{\,th}$        (GeV$$)   &   0.3          \\
\hline\hline
\end{tabular}\label{1s0_param}
\end{table}
\end{center}

\section{Discussion}\label{sect5}

\begin{figure}
\begin{center}
\begin{minipage}{0.49\textwidth}
\includegraphics[width=\textwidth]{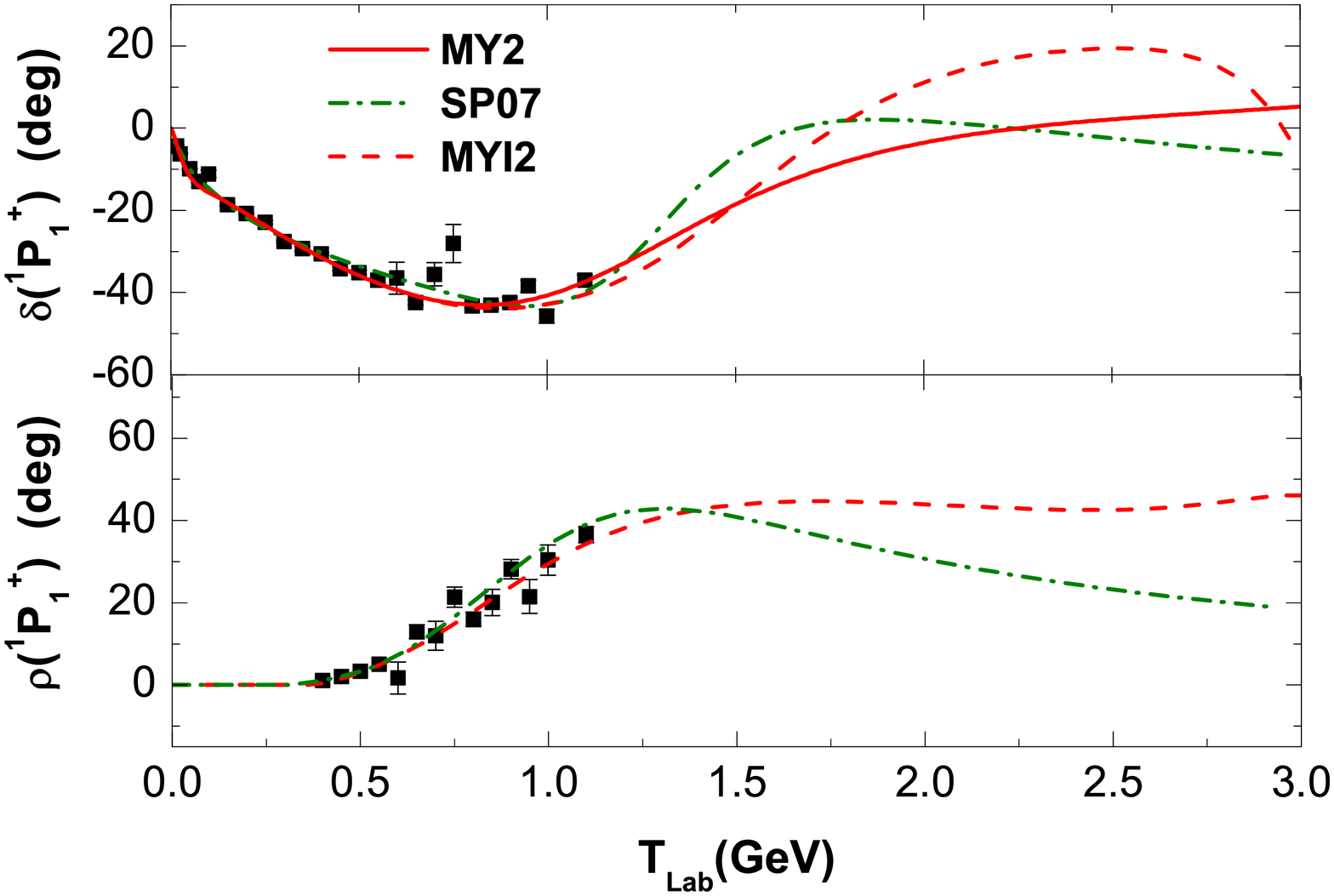}
\caption{{Phase shifts and inelasticity \protect\\ parameter for the $^1P_1^+$ partial-wave state.}}
\label{1p1}
\end{minipage}
\begin{minipage}{0.49\textwidth}
\includegraphics[width=\textwidth]{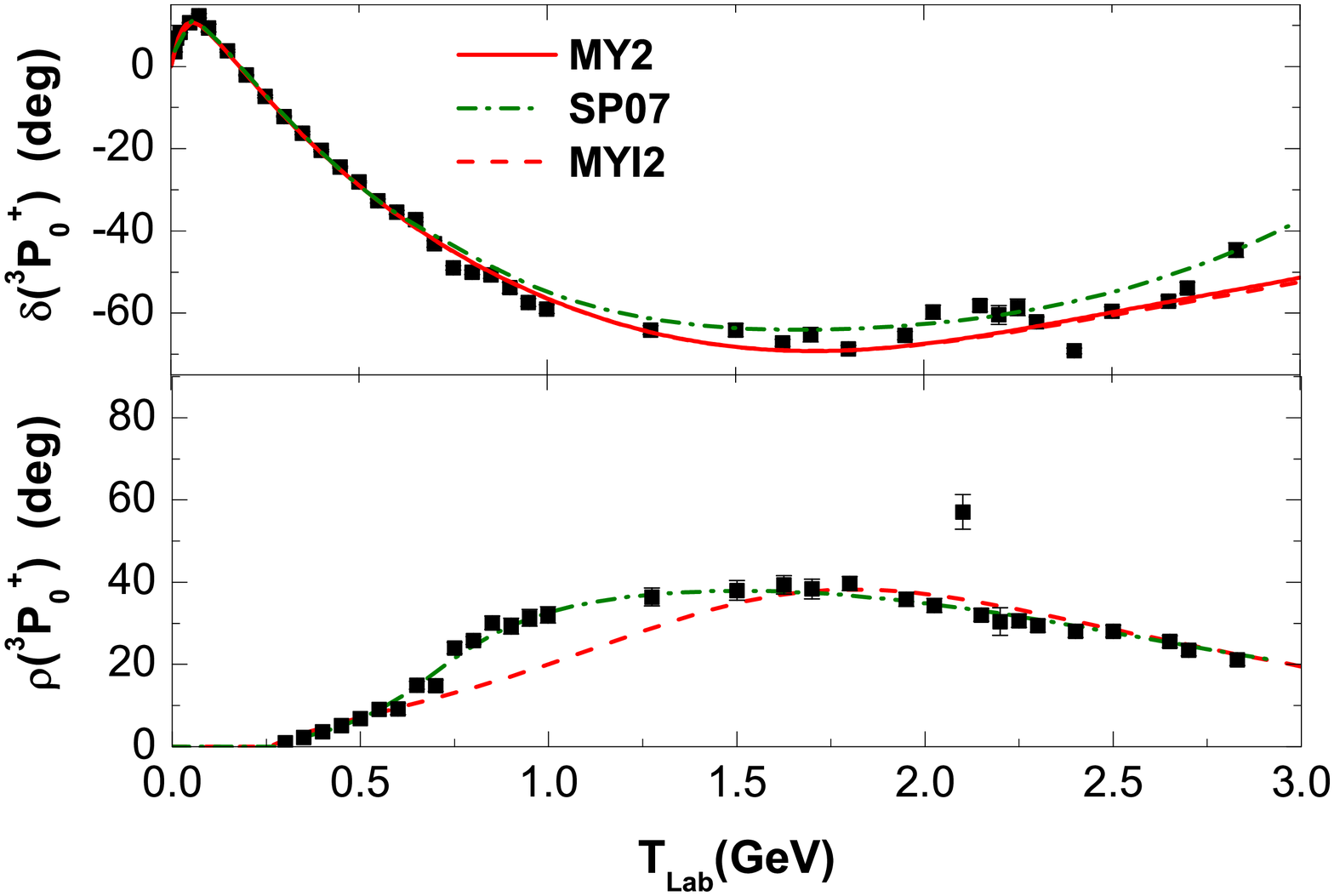}
\caption{{Phase shifts and inelasticity \protect\\ parameter for the $^3P_0^+$ partial-wave state.}}
\label{3p0}
\end{minipage}
\end{center}
\end{figure}

In. Fig.\ref{1p1}, we see that all calculations (MY2, MYI2, SP07)
give an excellent description of the phase shifts and two of them
(MYI2, SP07) - of the inelasticity parameter
for all available experimental data (up to $\tlab \sim 1.1$\,GeV)
for the $^1P_1^+$ partial-wave state. However, their
behavior is rather different at higher energies.
To make a choice in favor of one of them experimental
data in a wider energy range are necessary.

In Fig.\ref{3p0}, the results of the calculations for the $^3P_0^+$
partial-wave state are shown. All of them (MY2, MYI2, SP07)
demonstrate a reasonable agreement with the
experimental data for the phase shifts in the whole energy range
(up to $\tlab \sim 3$\,GeV).

\begin{figure}
\begin{center}
\begin{minipage}{0.8\textwidth}
\includegraphics[width=\textwidth]{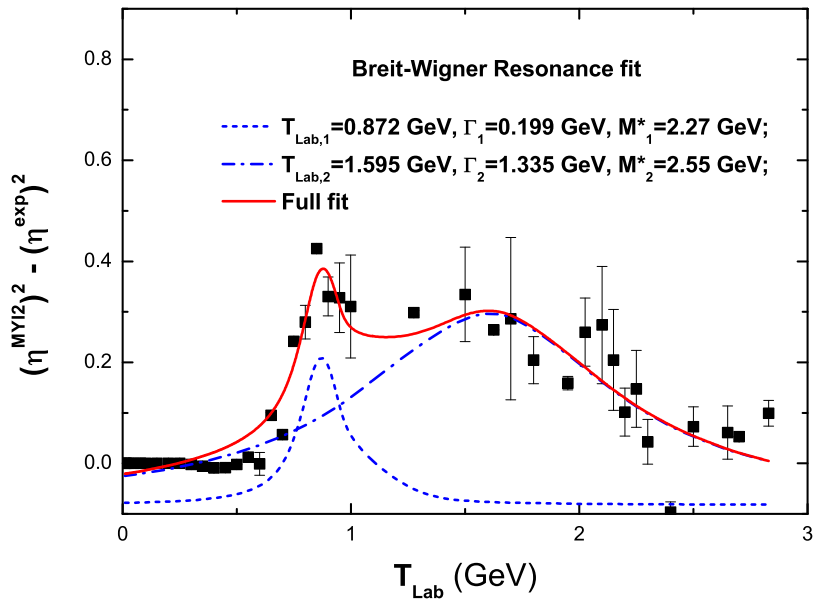}
\caption{Difference $\Delta\eta$ for the $^3P_0^+$ partial-wave state.}
\label{3P0_eta2}
\end{minipage}
\end{center}
\end{figure}

The description of the inelasticity parameter
is perfect for the MYI2 model and the SP07 solution up to $\tlab \sim 3$\,GeV
(except the energy interval $\tlab \sim 0.7$ - 1.4 \,GeV for MYI2).
The behavior of the inelasticity parameter for the $^3P_0^+$ state in this energy range
 needs a separate discussion. Let us consider the difference
\begin{equation}\label{diff}
\Delta\eta(\tlab)=(\eta^{\rm MYI2}(\tlab))^2-(\eta^{\rm exp}(\tlab))^2,
\end{equation}
Fig.\ref{3P0_eta2}, which is analyzed using the Breit-Wigner formula
\begin{equation}\label{BW}
\Delta\eta(\tlab) = C + \sum_{i=1,2}(2A_i/\pi)\,
\frac{\Gamma_i}{4(\tlab-M^*_i)^2 + \Gamma_i^2}
\end{equation}
where $A_i,C$ are constants, and $M_i,\Gamma_i$ are the effective
width and mass of the resonance systems, respectively.
It is seen that the obtained distribution is in perfect agreement with
$\Delta\eta$ in the considered energy range.
It may be interpreted
as a signal of two dybaryon resonances  with masses $M^*_1=2.27$ GeV,
$M^*_2=2.55$ GeV and widths $\Gamma_1 = 0.199$ GeV, $\Gamma_2 = 1.335$ GeV,
respectively. Of course, this fact needs a more careful analysis in future.

\begin{figure}
\begin{center}
\begin{minipage}{0.49\textwidth}
\includegraphics[width=\textwidth]{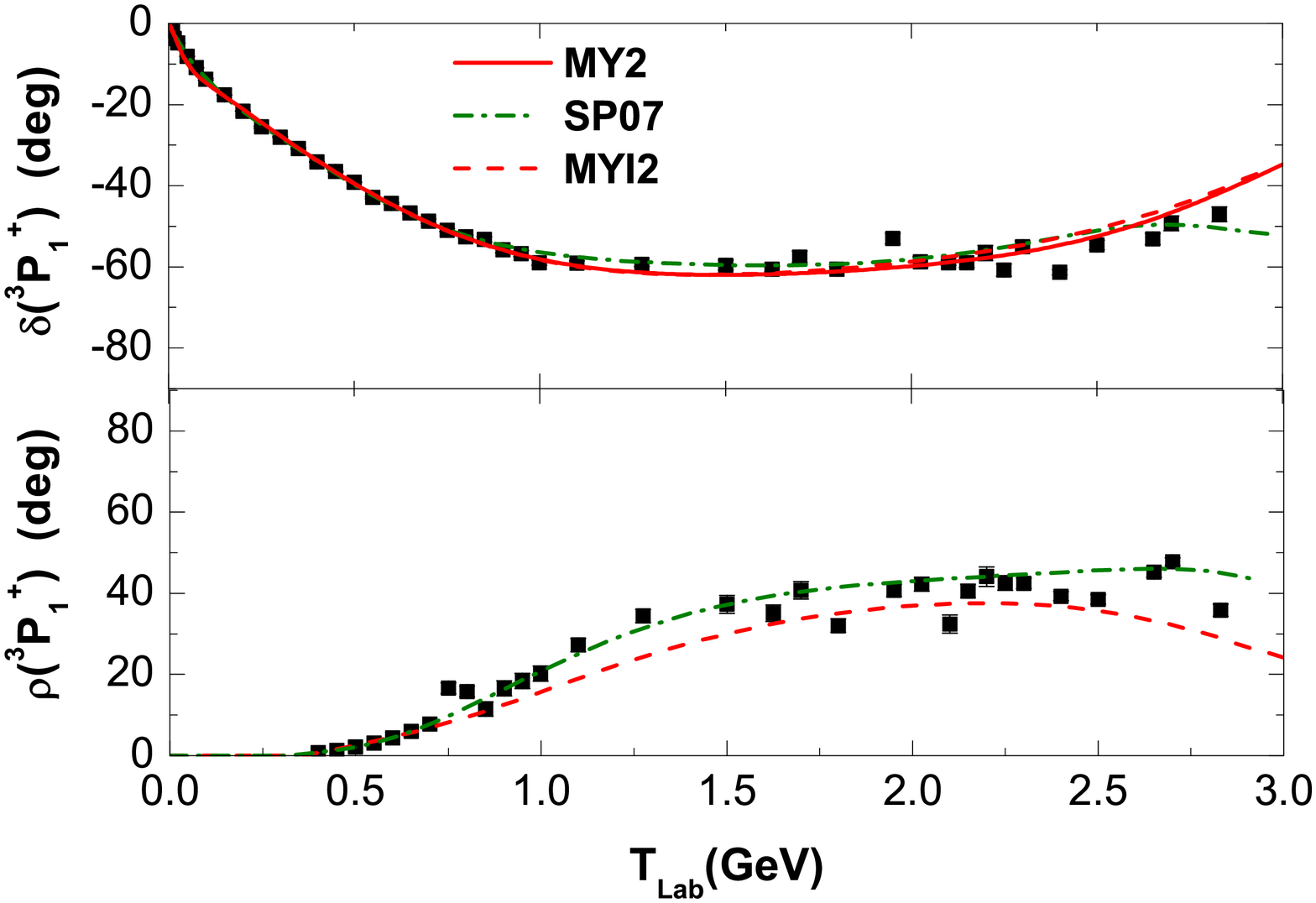}
\caption{{Phase shifts and inelasticity \protect\\ parameter for the $^3P_1^+$ partial-wave state.}}
\label{3p1}
\end{minipage}
\begin{minipage}{0.49\textwidth}
\includegraphics[width=\textwidth]{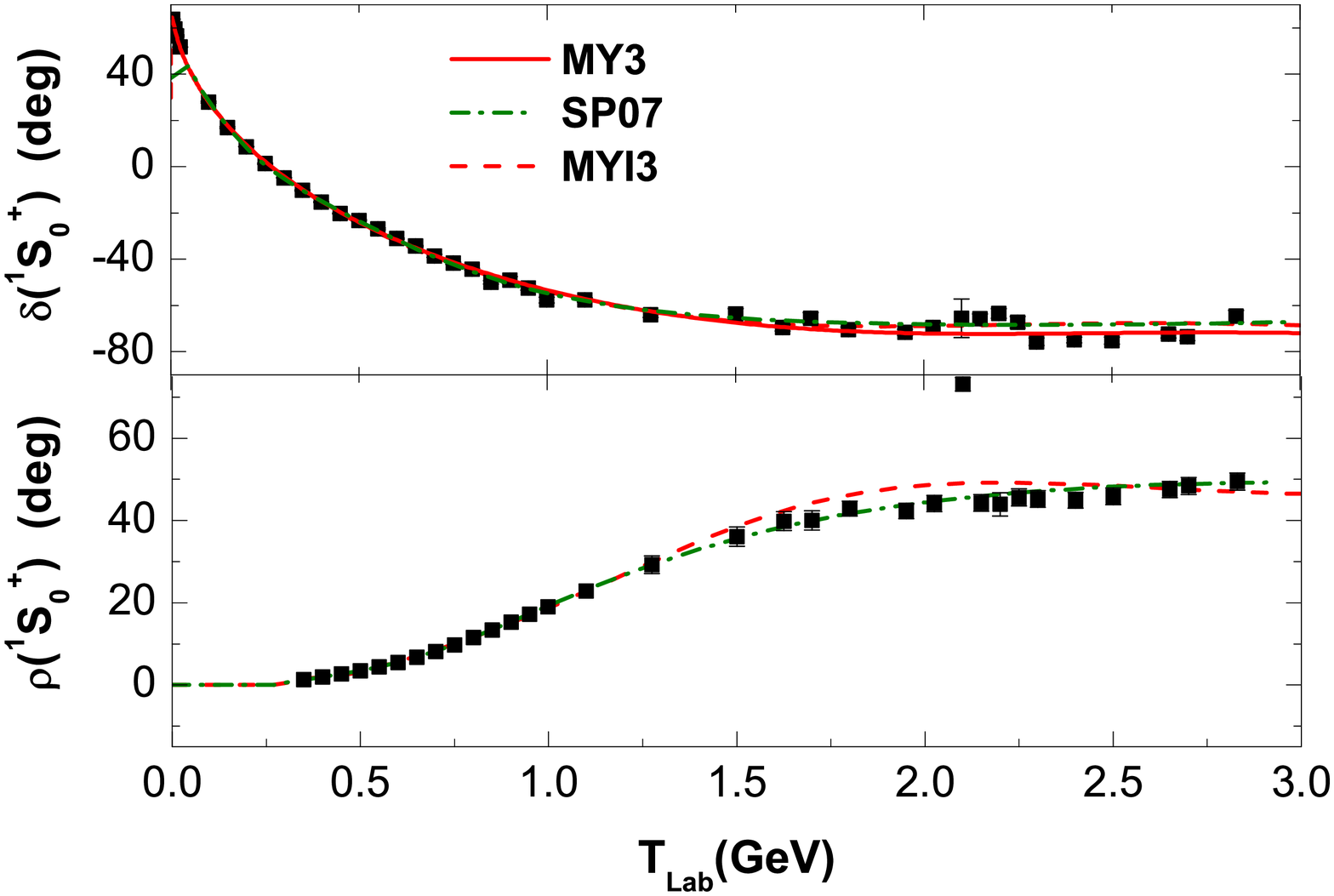}
\caption{{Phase shifts and inelasticity \protect\\ parameter for the $^1S_0^+$ partial-wave state.}}
\label{1s0}
\end{minipage}
\end{center}
\end{figure}

The phase shifts and the inelasticity parameter
for the $^3P_1^+$ partial-wave state are depicted in Fig.\ref{3p1}.
All results (MY2, MYI2, SP07) are acceptable
in the limits of the experimental errors
in the considered range of energies (up to $\tlab \sim 3$\,GeV).
However, MYI2 and SP07 give rather different description
for the inelasticity parameter.
Nevertheless, an uncertainty in the experimental data
values allows to accept both of them.

The $^1S_0^+$ partial-wave state is presented in Fig.\ref{1s0}.
It can be seen that all calculations show a perfect
agreement with the measured phase shifts in the whole energy range
(up to $\tlab \sim 3$\,GeV). The description of the inelasticity parameter
by the MYI3 potential and the SP07 solution is also good.

It is seen that the proposed MYIN potentials give a consistent description
of the existing experimental data for the phase shifts and the
inelasticity parameter.
It should be noted that since all parameters of the real separable interaction
kernel ($\lamr$, $\beta$ and $\alpha$) found in the previous analysis~\cite{npa1,npa2}
have been fixed, the phase shifts obtained using the MYN and MYIN models coincide
up to $\tlab < \tlab_{\,th}$ and are slightly different at $\tlab > \tlab_{\,th}$
for all considered partial-wave states except $^1P_1^+$.
The difference above $\tlab_{\,th}$ is explained by the influence
of the imaginary part $\lami$ in the MYIN potential
(see Eq.(\ref{lami})).


\section{Conclusion}\label{sect6}

The proposed complex potentials allow us to describe the inelasticity appearing
in the elastic $np$ scattering with increasing energy of the nucleons.
They have been constructed by the introduction of the imaginary part (minimal
extension) into the real-valued potentials elaborated earlier~\cite{npa1,npa2}.
In this case, the low-energy characteristics and the phase shifts below
the inelasticity threshold remain unchanged while above the threshold
the obtanied phase shifts slightly differ from the previous ones
(except the $^1P_1^+$ partial-wave state).

The imaginary part parameters have been found from the description of the
experimental data for the phase shifts and the inelasticity parameters
for the laboratory energy up to 3 GeV.

The deviation of the MYI2 curve from the experimental values
for the inelasticity parameter in the $^3P_0^+$ partial-wave state
can be interpreted as a presence of the dybaryon resonances.
However, this conclusion needs further investigation.

\section{Acknowledgments}\label{sect7}
We are grateful to Professor V. Karmanov for stimulating questions and useful
discussions.

\end{document}